\documentstyle[11pt]{article}

\begin{document}

\begin{flushright}
{BIHEP-TH-95-41}
\end{flushright}
\begin{center}

\renewcommand{\thefootnote}{\fnsymbol{footnote}}

{\LARGE 
Non-perturbative effect 
in the nucleon structure function and 
the Gottfried sum
\footnote{This work is supported by National Natural  
Science Foundation of China.}}
\vspace{5mm}

{\large Jian-Jun Yang$^{1,3}$, Bo-Qiang Ma$^{2,3}$, Guang-Lie Li$^{2,3}$} 

\vspace{5mm}
{\small 
$^1$Department of Physics, Nanjing Normal University, Nanjing 210024, China

$^2$CCAST(World Laboratory), P.O.Box 8730, Beijing 100080, China

$^3$Institute of High Energy Physics, Academia Sinica, P.O.Box 918(4), 
Beijing 100039, China\footnote{Mailing address. 
Email address: yangjj@bepc3.ihep.ac.cn}
}

\vspace{5mm}

{\large \bf Abstract } \\
\end{center}

We investigate the non-perturbative effect in the nucleon structure 
function and the Gottfried sum by using a non-perturbative quark
propagator with lowest dimensional condensate contributions
from the QCD vacuum. 
It is shown that the non-perturbative effect modifies the
conventional quark-parton model formula of the nucleon structure
function at finite $Q^2$ and suggests a non-trivial 
$Q^2$ dependence in the Gottfried sum.
    
\vspace{10mm}

\break

Within the framework of the quark parton model (QPM) \cite{parton}, 
deep inelastic lepton-nucleon  
scattering is viewed simply as the sum of elastic scatterings of the
lepton on ``free" 
quarks in the nucleon. 
In the derivation of the inelastic structure
function $F_{1,2}$, the incoherence assumption is made and the
initial and final state interactions are neglected.
The advantage of the 
simple QPM lies in the clear physical picture of the process and 
the simple form 
of the nucleon structure function which could be expressed as the sum of 
parton distributions weighted by the square of charge of 
corresponding partons (quarks plus antiquarks) for electron and muon
probes.
The QPM with QCD correction 
can give good descriptions 
of the large momentum transfer deep inelastic process. 
But recently, some experimental results are found to be beyond 
the predictions of the QPM. One of them is the violation of the
Gottfried sum rule from the New Muon Collaboration (NMC) measurement of
the ratio of the cross sections for unpolarized    
deep-inelastic scattering from deuterium and hydrogen targets
\cite{NMC91}.
The ratio
of structure functions $F_2^d(x,Q^2)/F_2^p(x,Q^2)$ is extracted in the range
$0.004<x<0.8$ at $Q^2=4.0$ GeV$^2$.
Assuming a smooth                        
extrapolation of the data for $F_{2}^{n}/F_{2}^{p}$ from $x=0.8$ to                 
$x=1$, adopting a Regge behavior $ax^{b}$ for                                      
$F_{2}^{p}-F_{2}^{n}$ (a flavor nonsinglet quantity) in the region                       
$x=0.004-0.15$ and then extrapolating it to $x=0$, the NMC reported            
the Gottfried sum, defined as 
\begin{equation}
S_G=\int_0^1\frac{{\rm d} x}{x}[F_2^p(x)-F_2^n(x)],
\end{equation}
with the result 
$S_{G}=0.235\pm0.026$.                                                            
This result is in striking contradiction with the Gottfried sum rule (GSR),
which 
would have $S_G=\frac{1}{3}$ in the QPM \cite{GSR}. 

Several different effects have been proposed as sources for this 
discrepancy between the NMC data and the GSR. They can be mainly 
referred to possibilities as follows: (1) a flavor distribution asymmetry
in the sea of nucleons \cite{Pre91,Pi}, 
i.e., an excess of $d\bar{d}$ over $u\bar{u}$
pairs in the proton due to the pauli exclusive principle and the excess 
of a u valence quark in the proton; (2) the unjustified $x\rightarrow0$
extrapolation of the available data \cite{Mar90};
(3) the nuclear                                                                     
effects like mesonic exchanges in the deuteron \cite{Kap91}                 
and nuclear bindings \cite{Epe92};
(4) isospin symmetry breaking between the proton and the neutrons
\cite{Ma92}.

We investigate in this letter the non-perturbative effect in
the nucleon structure function and  
the GSR.  
We first study the $<\bar{q}q>$-corrected quark propagator by taking
into account 
the lowest dimensional condensate contributions from the QCD vacuum.
Using the obtained non-perturbative quark propagator, 
we describe the non-perturbative effect 
in the nucleon structure function based on the QPM and
then discuss the consequence in  the GSR. 
A measurable non-trivial $Q^2$ dependence in the Gottfried sum is suggested.

Let us start by writing the free quark propagator 
\begin{equation}
i[S_F(x-y)]_{\alpha\beta}^{ab}
=<0|Tq_{\alpha}^{a}(x)\bar{q}_{\beta}^{b}(y)|0>,
\end{equation}
where $a$, $b$ are color indices 
and $\alpha$, $\beta$ are spinor indices. As a simple model, 
we consider, as shown in Fig.~1,  the lowest-order Feynman diagrams 
generating the dimension-3
quark condensate and dimension-4 gluon condensate contributions from
the QCD vacuum to the 
quark propagator. The free quark propagator without any modification of 
condensation can be expressed in momentum space as
\begin{equation}
S_F^{-1}(p)={\not \! p}-m_c 
\end{equation}
with the perturbative (current-) quark mass $m_c$ which can be neglected 
in large momentum transfer process. But in medium energy region, the
quark propagator should be modified by taking into account 
non-perturbative effects \cite{QCDSR,Non-P1,Non-P2}. In 
this letter 
we consider the non-perturbative effect from 
the Feynman diagrams as shown in Fig.~1 
to the quark propagator.
To  derive 
the effect of dimension-3 quark condensate contribution to the quark  
propagator, we use the nonperturbative vacuum expectation value (VEV) 
of two quark fields 
\begin{equation} 
<0|\bar{q}_\alpha^{a}(x)q_\beta^{b}(y)|0>_{NP}=\frac{1}{12}\delta_{ab}
 (1+\frac{im{\gamma}_{\mu}(x-y)^{\mu}}{4})_{\alpha\beta}<\bar{q}q>+\cdots
\;.
\end{equation}
For the dimension-4 gluon condensate contribution to the  
quark propagator, we take the nonperturbative VEV of two gluon fields
\begin{equation}
\begin{array}{clcr}
<0|A_\mu^a(x)A_\nu ^b(y)|0>_{NP}=\frac{1}{4}x^{\lambda}y^{\rho}
<0|G_{\lambda\mu}^aG_{\rho\nu}^b|0>+\cdot\cdot\cdot  \\
=\frac{1}{4}\frac{1}{96}x^{\lambda}y^{\rho}
(g_{\lambda\rho}g_{\mu\nu}-g_{\lambda\nu}g_{\mu\rho})
<GG>+\cdot\cdot\cdot
\end{array}
\end{equation}
in the fixed point gauge $x^{\mu}A_{\mu}(x)=0$,
where 
\begin{equation}  
  <GG>=<0|G_{\lambda\mu}^a(0)G_{\lambda\mu}^a(0)|0>.
\end{equation}
Under the chain approximation, one can obtain the complete quark propagator 
in momentum space \cite{Non-P2}
\begin{equation}
\begin{array}{clcr}
S_F^{-1}(p)={\not \! p}[1+\frac{g_s^2<\bar{q}q>(1-\xi)m}
{9p^4}+\frac{g_s^2<GG>m_c^2}{12(p^2-m_c^2)^3}]
\\
-[m_c+\frac{g_s^2<\bar{q}q>(4-\xi)}
{9p^2}+\frac{g_s^2<GG>m_cp^2}{12(p^2-m_c^2)^3}],
\end{array}
\end{equation}
where $m$ in $S_F^{-1}(p)$ arises 
from incorporating the QCD equation of motion 
\begin{equation}
(i{\not \! \! D} -m)\psi=0.
\end{equation}
It is necessary to emphasize that $m$, which includes 
the effect of the condensates 
of non-perturbative QCD, is different from 
the purely perturbative (current-) 
quark mass $m_c$. 
Note also that $S_F^{-1}(p)$ is gauge parameter $\xi$ dependent
due to the internal gluon line appearing in Fig.~1(b).
In common sense, the current  
quark mass $m_c$ is small, 
and it can be neglected in large momentum transfer process,
which is equivalent to 
neglecting the gluon condensate term in $S_F^{-1}(p)$.
Therefore $S_F^{-1}(p)$ can be rewritten as 
\begin{equation} 
S_F^{-1}(p)={\not \! p}-M(p)
\end{equation}
with
\begin{equation}
M(p)=\frac{g_s^2<\bar{q}q>}{9p^2}[(4-\xi)-\frac{(1-\xi){\not \! p}m}{p^2}].
\end{equation}

We require the pole of the $<\bar{q}q>$-corrected 
propagator corresponding   
with $m$ in equation of motion (8), i.e.,  
\begin{equation}
M(p)|_{{\not  p}=m}=\frac{g_s^2<\bar{q}q>}{3m^2}=m.
\end{equation}
From this equation, one can obtain the solution of $m$ 
which is independent of gauge parameter
$\xi$ 
\begin{equation}
m=M(p)|_{{\not  p}=m}=(\frac{4\pi\alpha_s(Q^2)<\bar{q}q>}3)^{1/3}.
\end{equation}
Thus the $<\bar{q}q>$-corrected quark propagator can be written as  
\begin{equation} 
S_F^{-1}(p)={\not \! p}-(\frac{4\pi\alpha_s(Q^2)<\bar{q}q>}3)^{1/3}, 
\label{eq:qp}
\end{equation}
where the strong coupling constant 
$\alpha_s(Q^2)=\frac{4\pi}{\beta_0ln\frac{Q^2}{\Lambda^2}}$
with $\beta_0=11-\frac23N_F$ and the dimensional parameter
$\Lambda=0.250$ GeV for
$N_F=3$.

We now discuss the effect of the $<\bar{q}q>$ condensate in the 
nucleon structure function by means of the non-perturbative quark propagator
Eq.~(\ref{eq:qp}).
Consider the inclusive lepton-nucleon scattering
\begin{equation}
l+N \rightarrow l+X
\end{equation}
where the hadronic structure is entirely contained in the tensor $W_{\mu\nu}$
\cite{HKS} 
\begin{equation}
\begin{array}{clcr}
W_{\mu\nu}=(2\pi)^3\sum\limits_{X}
<P|J_{\mu}|X><X|J_{\nu}|P>\delta^4(P_X-P-q) 
\\ 
=(-g_{\mu\nu}+\frac{q_{\mu}q_{\nu}}{q^2})W_1
+\frac1{M^2}(P_{\mu}-\frac{P.q}{q^2}q_{\mu})
(P_{\nu}-\frac{P.q}{q^2}q_{\nu})W_2,
\end{array}
\end{equation}
here $M$ is the mass of the nucleon.
If $W_{\mu\nu}$ is given, one can extract $W_1$ and $W_2$ 
through the following formulas:
\begin{equation} 
W_1=\frac12[c_2-(1-\frac{\nu^2}{q^2})c_1](1-\frac{\nu^2}{q^2})^{-1};
\label{eq:w1}
\end{equation}
\begin{equation}
W_2=\frac12[3c_2-(1-\frac{\nu^2}{q^2})c_1](1-\frac{\nu^2}{q^2})^{-2},
\label{eq:w2}
\end{equation}
with $c_1\equiv W_{{ } \mu}^\mu$ and
$c_2\equiv \frac{P^{\mu} P^{\nu}}{M^2}W_{\mu\nu}$ \cite{HKS}.
All non-perturbative effects are entirely contained in $W_{\mu\nu}$. 
In this letter, we try to study the non-perturbative effect 
in the nucleon structure function from the quarks 
in the QCD physical vacuum.
We suppose that the proton is made up of bound 
partons that appear as ``free" Dirac particles 
but 
with the non-perturbative 
propagator because of being in the QCD vacuum.
With the incoherence assumption,
one parton contribution to $W_{\mu\nu}$ is  
\begin{equation}
\begin{array}{clcr}
w_{\mu\nu}=(2\pi)^3\frac12\sum\limits_{s,s^{\prime}}
\sum\limits_{p^\prime}<\vec{p},s|J_{\mu}|
\vec{p^{\prime}},s^{\prime}><\vec{p^\prime},s^{\prime}|J_{\nu}|\vec{p},s>
\delta^4(p^\prime-p-q)
\\
=e_i^2\int 
\frac{d^3p^{\prime}}{2p^{\prime}_0}\delta^4(p^{\prime}-p-q)
\frac12 {\rm Tr}\,[\gamma_{\mu}({\not \! p}^{\prime}+m)\gamma_{\nu}
({\not \! p}+m)].
\end{array}
\end{equation}
According to the trace theorem that the trace of an odd number of $\gamma's$ vanishes,
$w_{\mu\nu}$ can  also be equivalently expressed as
\begin{equation} 
 w_{\mu\nu}=e_i^2\int 
\frac{d^3p^{\prime}}{2p^{\prime}_0}\delta^4(p^{\prime}-p-q)
\frac12 {\rm Tr}\, [\gamma_{\mu}({\not \! p}^{\prime}-m)\gamma_{\nu}
({\not \! p}-m)];
\end{equation}
i.e.,
\begin{equation} 
w_{\mu\nu}=e_i^2\int \frac{d^3p^{\prime}}
{2p^{\prime}_0}\delta^4(p^{\prime}-p-q)
\frac12 {\rm Tr}\,
[\gamma_{\mu}{S_F^{-1}(p^{\prime})}\gamma_{\nu}{S_F^{-1}(p)}],
\end{equation}
\noindent
where $iS_F(p)$ is the quark propagator and $e_i$ is the charge of quark
in unit of $e$. 
Generally, one should take the complete non-perturbative quark propagator
including the correction  due to $<\bar{q}q>$,
$<GG>$ and higher dimensional condensate. However, as a simple qualitative
analysis,
we take only the $<\bar{q}q>$-corrected quark propagator given 
by Eq.~(\ref{eq:qp}).
For the sake of simplicity,
we adopt the parton picture in which the parton 4-momentum is expressed as 
$p^\mu=yP^\mu$ $(0\leq y\leq 1)$ with the nucleon 4-momentum $P^\mu$; 
i.e., we 
assume that all transverse momenta are negligible
and that no parton moves oppositely to the nucleon.
Using Eqs.~(\ref{eq:w1})  
and (\ref{eq:w2}), we can extract one quark contribution of type $i$
quark 
to $W_2$, and the corresponding
contribution to the nucleon structure function  $F_2^{(i)}=\nu w_2$ is 
\begin{equation}
F_2^{(i)}(y)=2Mx^2e_i^2\delta(y-x)R_{NP}(Q^2),
\label{eq:Fi}
\end{equation}
with
\begin{equation} 
x=\frac{Q^2}{2M\nu},
\end{equation}
and
\begin{equation}
R_{NP}(Q^2)=1-\frac{4}{Q^2}(\frac{4\pi\alpha_s<\bar{q}q>}{3})^{2/3}.
\end{equation}
Suppose that the
nucleon 
state contains $f_i(y){\rm d} y $
parton states of the type $i$ in the interval ${\rm d} y$, then 
\begin{equation}
F_{2}=\sum\limits_{i}\int_0^1{\rm d}yf_i(y)F_{2}^{(i)}.
\end{equation}
We adopt the convention of Ref.~\cite{HKS}
in which a parton state has $2p_0$ partons per unit volume,
while a nucleon state has $P_0/M$ nucleons per unit volume.
Therefore, in one nucleon, the number of
partons of type $i$, in the interval ${\rm d} y$ is $f_i(y)$ multiplied by 
$\frac{2p_0}{(P_0/M)}=2 M y$,
i.e., $q_i(y){\rm d} y=2 M y f_i(y){\rm d} y$, where $q_i(y)$ is
the quark parton distribution with  
constraint of parton flavor number conservation. 
Summing all contributions of all quarks in the nucleon,
we obtain the structure function of the nucleon
\begin{equation}
\begin{array}{clcr}
F_2(x)=\sum\limits_iq_i(x)xe_i^2R_{NP}(Q^2)
=\sum\limits_i\tilde{q}_i(x)xe_i^2,
\end{array}
\label{eq:F2}
\end{equation}
where
\begin{equation}
\tilde{q}_i(x)=q_i(x)R_{NP}(Q^2)
\end{equation}
which is different from $q_i(x)$ since $q_i(x)$ represents the 
probability distribution of quarks of type $i$ and satisfies 
the parton
flavor number sum rule, but $\tilde{q}_i(x)$ does not.
From Eq.~(\ref{eq:F2}), we see that the structure function, with
a reducing factor of $R_{NP}(Q^2)$, 
is no longer simply the sum of the parton distributions
multiplied 
by the square charge 
of corresponding partons (quarks plus antiquarks) in the
nucleon at finite $Q^2$.
In fact, there are some results in recent 
experimental data which show 
explicit violations of conventional parton sum rules. In this letter
we take  
the violation of GSR 
as an example, although the non-perturbative
effect should also have consequences in other parton sums. 
To show explicitly the non-perturbative effect in the 
structure function, 
we give the values of $R_{NP}(Q^2)$ for corresponding $Q^2$ in Tab.~1.
In the estimate of $R_{NP}(Q^2)$, we have taken 
the standard phenomenological 
value of the quark condensate 
$<\bar{q}q>=-(0.25{\rm GeV})^3$ obtained from QCD sum
rule \cite{QCDSR}.

The available observed Gottfried sum is $S_G=0.235$ 
at $Q^2=4$ GeV$^2$ by the NMC
experiment and we need $R_{NP}=0.705$ to explain the data if there is
no any other sources for the violation of the GSR. Our calculated value
is $R_{NP}=0.922$ at $Q^2=4$ GeV$^2$ and is
too small to explain the data. However, we see from Tab.~1 that
there is non-trivial $Q^2$ dependence in the calculated $R_{NP}(Q^2)$.
Thus we would expect a measurable $Q^2$ dependence in the Gottfried
sum. From the observation of $Q^2$ dependences in several other parton 
sums, we know that this expectation is not unreasonable and the
magnitude of the $Q^2$ dependence is of the order $0.06$GeV$^2/Q^2$ which
is consistent with the $Q^2$ dependence as observed in experiments
and as estimated from the perturbative
QCD (pQCD) and higher twist effects in other parton sums 
\cite{Man95}. 
The pQCD corrections to flavor number conservation
are expected to be small and to have little consequence
on the GSR \cite{Ros79}.
The effect considered in this letter is from the
non-perturbative QCD vacuum and thus should include both of some 
pQCD and higher twist effects.     
  
In summary, we investigate the non-perturbative effect in the nucleon 
structure function and the Gottfried sum by taking into account
the lowest dimensional condensate contributions from the QCD
vacuum in the quark propagator.
We find a non-trivial modification of the conventional quark parton
model formula of the nucleon structure function at finite $Q^2$ 
and suggest a
measurable $Q^2$ dependence in the Gottfried sum.


\break
\begin{center}
Table 1. $R_{NP}(Q^2)$ for corresponding $Q^2$\\
\vskip 0.5cm
\begin{tabular}{|c|c|}\hline
$Q^2$&$R_{NP}(Q^2)$\\ \hline
2.0&0.823\\ \hline
4.0&0.922\\ \hline
10&0.973\\ \hline
20&0.987\\ \hline
100&0.998\\ \hline
\end{tabular}
\end{center}
                                                                                
\break                                                                          
\noindent                                                                       
{\large \bf Figure Captions}                                                    
\renewcommand{\theenumi}{\ Fig.~\arabic{enumi}}                                 
\begin{enumerate}                                                               
\item                                                                           
The non-perturbative quark propagator including:\\ 
(a). The perturbative free quark propagator;\\
(b). Lowest-order correction due to the nonvanishing  
value of  $<\bar{q}q>$;\\
(c). Lowest-order correction due to the nonvanishing  
value of  $<GG>$.
\end{enumerate}                                                                 
                                                                                

\newpage                                                                        
                                                                                
\begin{thebibliography}{99}                                                     
                                                                                
\bibitem{parton}
      R.~P.~Feynman,
      Phys. Rev. Lett. {\bf 23} (1969) 1415;

      J.~D.~Bjorken, Phys.~Rev.~{\bf 179} (1969) 1547;

      J.~D.~Bjorken and E.~A.~Paschos, 
      Phys.~Rev.~{\bf 185} (1969) 1975.  

\bibitem{NMC91} NM Collab., P.~Amaudruz {\it et al.},                           
                Phys.~Rev.~Lett. {\bf 66} (1991) 2712;
                M.~Arneodo {\it et al.}, 
                Phys.~Rev.~{\bf D 50} (1994) R1.                             
 
\bibitem{GSR}   K.~Gottfried,                                                   
                Phys.~Rev.~Lett. {\bf 18} (1967) 1174.                          
                                                                                
\bibitem{Pre91} G.~Preparata, P.~G.~Ratcliffe, and J.~Soffer,                   
                Phys.~Rev.~Lett. {\bf 66} (1991) 687.                           
                                                                                
\bibitem{Pi}    E.~M.~Henley and G.~A.~Miller,                                  
                Phys.~Lett.~{\bf B 251} (1990) 453;                             
                                                                                
                S.~Kumano, Phys.~Rev.~{\bf D 43} (1991) 59;                     
                {\bf D 43} (1991) 3067;                                         
                                                                                
                A.~Signal, A.~W.~Schreiber, and A.~W.~Thomas,                   
                Mod.~Phys.~Lett.~{\bf A 6} (1991) 271.                          
                                                                                
                                                                               
\bibitem{Mar90} A.~D.~Martin, W.~J.~Stirling, and R.~G.~Roberts,                
                Phys.~Lett.~{\bf B 252} (1990) 653;                             
                                                                                
                S.~D.~Ellis and W.~J.~Stirling,                                 
                Phys.~Lett.~{\bf B 256} (1991) 258.                             
                                                                                
\bibitem{Kap91} L.~P.~Kaptari and A.~Yu.~Umnikov,                               
                Phys.~Lett.~{\bf B 272} (1991) 359.                             
                                                                                
\bibitem{Epe92} L.~N.~Epele, H.~Fanchiotti, C.~A.~Carc\'ia Canal,               
                and R.~Sassot,                                                  
                Phys.~Lett.~{\bf B 275} (1992) 155.                             
 
\bibitem{Ma92}  B.~-Q.~Ma, Phys.~Lett.~{\bf B 274} (1992) 111;                  
 
                B.~-Q.~Ma, A.~Sch\"afer, and W.~Greiner,             
                Phys.~Rev.~{\bf D 47} (1993) 51; J.~Phys.~{G 20}
                (1994) 719.                                

\bibitem{QCDSR}
         M.A.~Shifman, A.I.~Vainshtein and V.I.~Zakharov, Nucl.~Phys.
	 {\bf B147} (1979) 385; 448.

\bibitem{Non-P1}
         T.~I.~Larsson, Phys. Rev. {\bf D32} (1985) 956;

         T.~Huang and Z.~Huang, Phys.~Rev. {\bf D39} (1988)1213;

         L.~J.~Reinders, H.~Rubinstein and S.~Yazaki, Phys. Rep. {\bf
         127} (1985) 1.

\bibitem{Non-P2}
         V.~Elias and T.~G.~Steele and M.~D.~Scadron, Phys. Rev. {\bf
         D38} (1988) 1584.

\bibitem{HKS}
         Kerson Huang, {\it Quarks, Leptons \& Gauge Fields} 
         (World Scientific, 1982) p.35.                                     

\bibitem{Man95}
         See, e.g., L.~Mankiewicz, E.~Stein, and A.~Sch\"afer,
                Preprint hep-ph/9510418.                               

\bibitem{Ros79}
         D.~R.~Ross and C.~T.~Sachrajda, Nucl.~Phys.~{\bf B 149}
         (1979) 497.

\end{thebibliography}
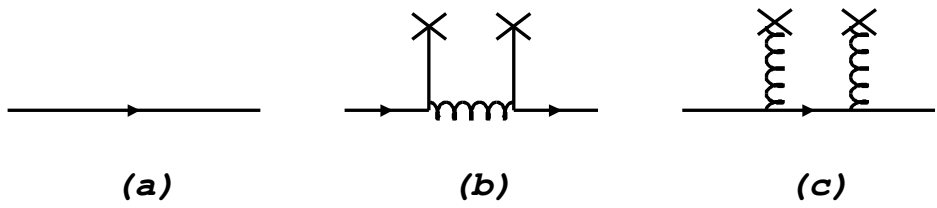
\end{document}